\pdfoutput=1
\documentclass[aps,prl,reprint,english,showpacs,showkeys,preprintnumbers,superscriptaddress]{revtex4-1}

\usepackage[T1]{fontenc}
\usepackage[latin9]{inputenc}
\usepackage{xcolor}
\usepackage{pdfcolmk}
\usepackage{babel}
\usepackage{amstext}
\usepackage{graphicx}
\usepackage{esint}
\PassOptionsToPackage{normalem}{ulem}
\usepackage{ulem}
\usepackage[unicode=true,pdfusetitle,
 bookmarks=true,bookmarksnumbered=false,bookmarksopen=false,
 breaklinks=false,pdfborder={0 0 0},pdfborderstyle={},backref=false,colorlinks=false]
 {hyperref}

\makeatletter

\providecolor{lyxadded}{rgb}{0,0,1}
\providecolor{lyxdeleted}{rgb}{1,0,0}

\DeclareRobustCommand{\lyxsout}[1]{\ifx\\#1\else\sout{#1}\fi}

\makeatother

\begin{document}

\title{Jet shape modifications in holographic dijet systems}

 \author{Jasmine Brewer} 
\email{jtbrewer@mit.edu} 
\affiliation{Center for Theoretical Physics, MIT, Cambridge, MA 02139, USA}
\author{Andrey Sadofyev} 
\affiliation{Theoretical Division, MS B283, Los Alamos National Laboratory, Los Alamos, NM 87545, USA}
\author{Wilke van der Schee} 
\email{w.vanderschee@uu.nl} 
\affiliation{Center for Theoretical Physics, MIT, Cambridge, MA 02139, USA}
\affiliation{Institute for Theoretical Physics and Center for Extreme Matter and Emergent Phenomena, Utrecht University, Leuvenlaan 4, 3584 CE Utrecht, The Netherlands}





\preprint{MIT-CTP/5070}
\begin{abstract}
We present a coherent model that combines jet production from perturbative
QCD with strongly-coupled jet-medium interactions described in holography.
We use this model to study the modification of an ensemble of jets
upon propagation through a quark-gluon plasma resembling central
heavy ion collisions. Here the modification
of the dijet asymmetry depends strongly on the subleading jet width,
which can therefore be an important observable for studying jet-medium
interactions. We furthermore show that the modification of the shape
of the leading jet is relatively insensitive to the dijet asymmetry,
whereas the subleading jet shape modification is much larger for more
imbalanced dijets.
Finally, we compare the results of our holographic model to a recent CMS measurement.
\end{abstract}
\maketitle
\noindent \textbf{Introduction - }The discovery that the quark-gluon
plasma (QGP) produced in heavy ion collisions (HIC) at RHIC and the
LHC is strongly coupled makes it both interesting and difficult to
understand \cite{CasalderreySolana:2011us,Busza:2018rrf}. One excellent
way to probe this plasma is to make use of highly energetic quarks
and gluons that are naturally produced during the collision and study
their properties after they traverse the medium \cite{Connors:2017ptx}.
The production of such energetic sprays of particles, called jets,
in elementary collisions is well understood within a perturbative
QCD (pQCD) framework, but the interaction of these jets with QGP is
theoretically more challenging since the physics of the plasma itself
is strongly coupled. Here we will start with an ensemble of jets where
the energies, widths and dijet asymmetry of the produced jets are
fixed from pQCD computations, but the subsequent evolution through
QGP is fully determined from holography \cite{Rajagopal:2016uip,Brewer:2017fqy}.

One of the most famous observables showing the strongly interacting
nature of the QGP is the increased imbalance of jet energies in a
dijet system. This is characterized by the modification of the dijet
asymmetry $A_{J}=(p_{1}-p_{2})/(p_{1}+p_{2})$, with $p_{1,2}$ the
transverse momentum of the leading and subleading jet.
The energy
loss of jets depends on the path length through the medium, but even
for a constant path length the energy loss fluctuates from jet to
jet. Recent work found that solely including fluctuations in the energy
loss distribution of centrally-produced dijets can explain the dijet
asymmetry observed in HIC, even though in this case both jets have
equal path lengths \cite{Milhano:2015mng}. 
Indeed, the fluctuations in energy loss for fixed path lengths can be comparable to the energy loss itself \cite{Escobedo:2016jbm}.

We confirm these studies by showing that the dijet asymmetry in our model
is insensitive to the starting position of the dijet, provided that
the jet widths of each jet fluctuate independently (in our model the
jet width is crucial for the energy loss). In the theoretical case
where both jets are restricted to have the same jet width we find
that the position and hence the path length imbalance is important
to modify the dijet asymmetry. Curiously, either the path length variation
or the energy loss fluctuation results in an almost equal dijet asymmetry
distribution. However, a coherent picture with appropriate nuclear
modification factor and jet shape modifications can only be obtained
when both effects are included. 

The width of a jet, the energy loss and the dijet asymmetry have an interesting interplay in holography. Every jet widens and wider jets lose more energy \cite{Chesler:2015nqz, Rajagopal:2016uip}. The steeply falling spectrum of jets together with the $p_T$ cut can however act as a selection mechanism for narrower jets, which lose less energy. We will see that this mechanism is especially relevant for leading jets, but for subleading jets the energy loss simply moves the dijet to a more asymmetric dijet bin, since the subleading jet already passes the $p_T$ cut by virtue of the leading jet $p_T$.

Our analysis suggests two new observables that can be particularly
informative for the interaction of jets and QGP. We first highlight
the dependence of the dijet asymmetry on the subleading jet width
and show that wide subleading jets lead to imbalanced dijets. Secondly,
we show the jet shape modifications of subleading and leading jets
binned for different values of the dijet asymmetry. Our model suggests
that the leading jet is strongly modified independent of $A_{J}$,
whereas the most imbalanced dijets have the widest subleading jets.
This observation matches qualitatively with simulations done in \textsc{Jewel},
but the effect of the subleading jet shapes on the dijet asymmetry
is stronger in our holographic model. For related recent works on the jet substructure see also \cite{Chien:2015hda,Casalderrey-Solana:2016jvj,KunnawalkamElayavalli:2017hxo,Luo:2018pto, He:2018xjv, Pablos:2019ngg, Casalderrey-Solana:2019ubu, Chang:2019sae, Du:2020pmp, Ke:2020clc, Vaidya:2020lih, Sadofyev:2021ohn} and references therein. 

\begin{figure*}[t]
\begin{centering}
\includegraphics[width=6cm]{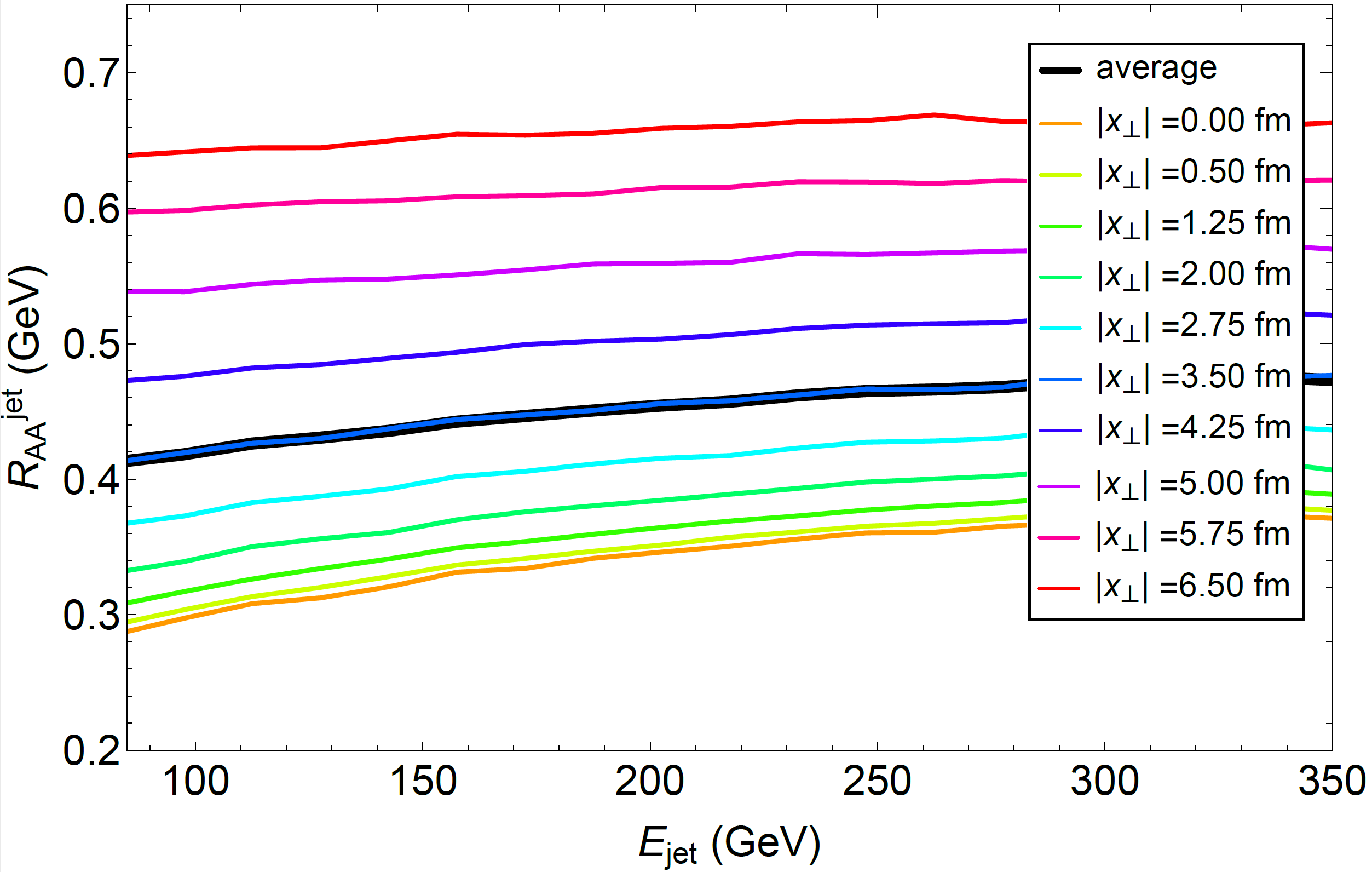}\includegraphics[width=6cm]{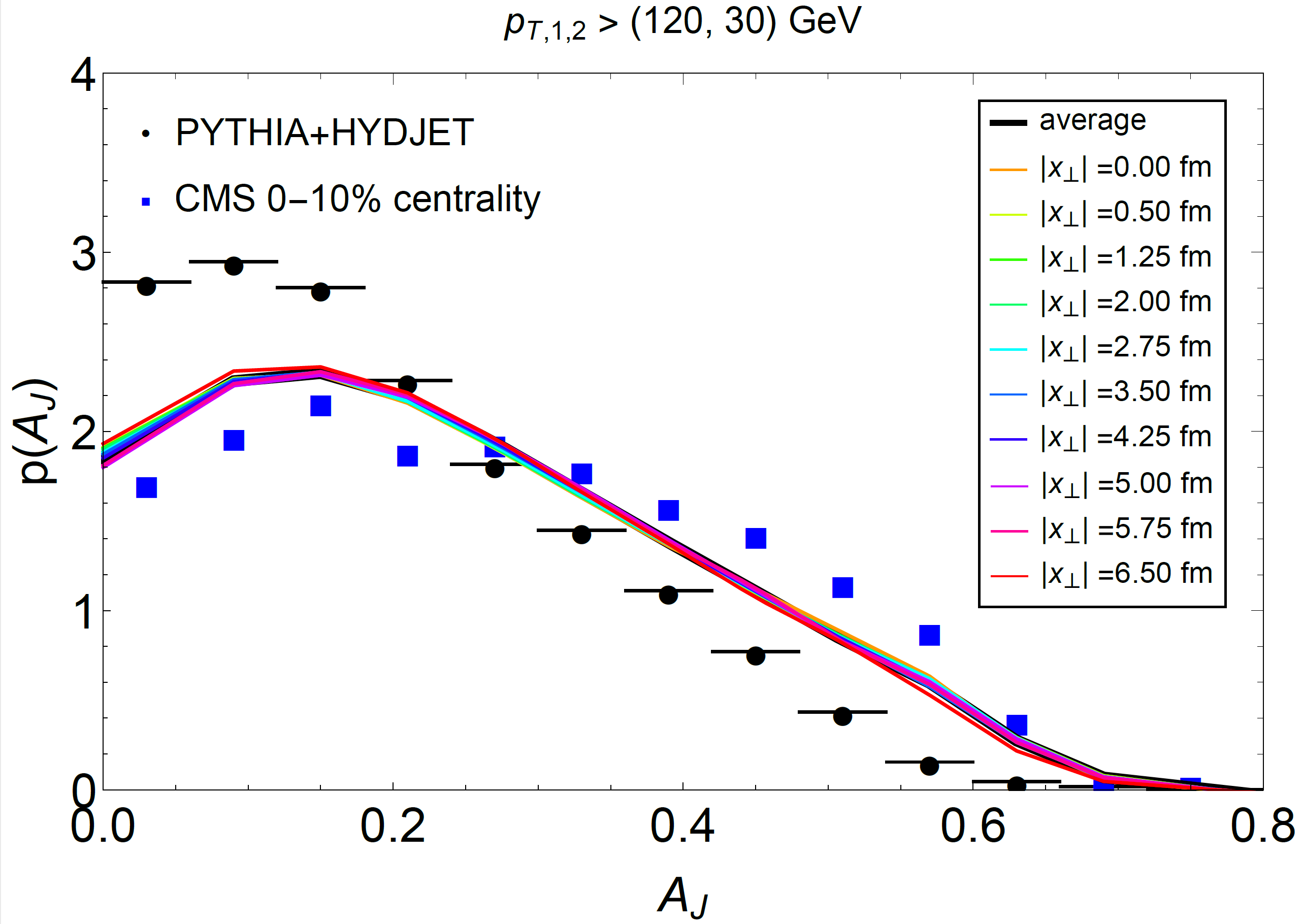}\includegraphics[width=6cm]{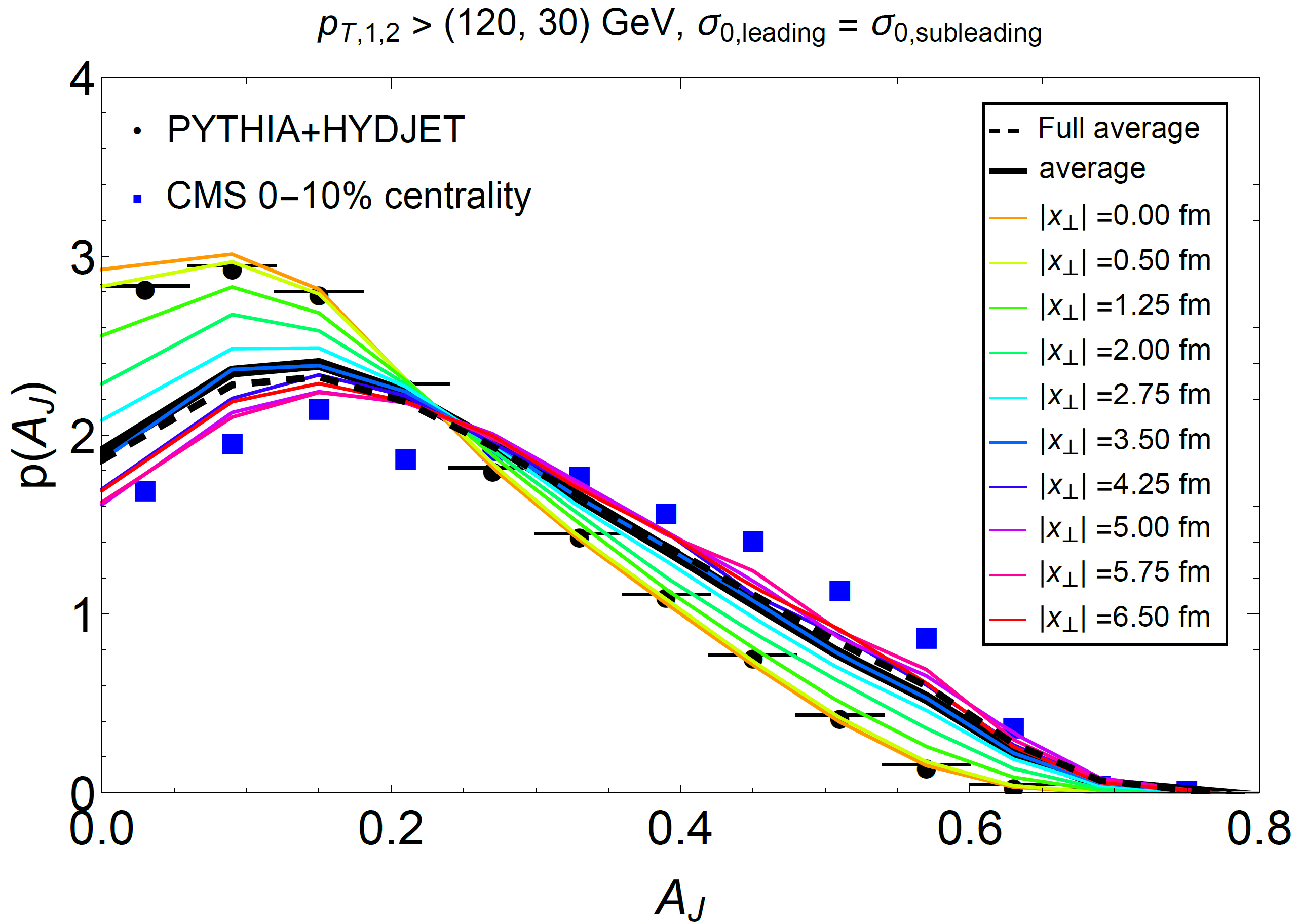}
\par\end{centering}
\caption{(left) Jet nuclear modification factor $R_{AA}^{\text{jet}}$ for
different starting positions of jets in the transverse plane and the
physical value (black). (middle) The corresponding dijet asymmetry
distribution is surprisingly insensitive to the starting position,
even though central starting positions have balanced path lengths
\cite{Milhano:2015mng}. Jets produced at the center do lose more
energy (as evident from the $R_{AA}^{jet}$), and are also more heavily
modified (see also Fig. \ref{fig:differentiated}). (right) When (artificially)
demanding that both jets in a dijet system have the same jet width
the dijet distribution does depend on the path length imbalance. Interestingly
the average dijet imbalance for jets with the same width (solid black)
is almost equal to the physical case where both jet widths fluctuate
independently (dashed black). The overlayed data points show the input
$A_{J}$ distribution (black circles) from \textsc{Pythia}+\textsc{Hydjet} simulations
and the modified distribution (blue squares) measured by CMS in \cite{Chatrchyan:2012nia}.
\label{fig:RAA}}
\end{figure*}

The aim of this work is to qualitatively study the effects of energy
loss from holography and to use those insights to identify observables
that capture the physics of jet-medium interactions. We do not attempt
to model all aspects of jet physics in the holographic calculation
and specifically do not include third jets, incoherent substructure
inside the subleading jet, hadronization or a jet finding algorithm.
All of these effects can have important contributions, particularly
on the subleading jet shape far from the jet axis as we discuss later.
Finally, this work has become especially relevant as CMS recently published results similar to our proposed measurement \cite{Sirunyan:2021jty}. In the discussion we comment on a comparison with our model predictions.

\noindent \textbf{The model - }In holographic models the propagation
of a quark-antiquark pair at large coupling is described by the dynamics
of a classical string in Anti-de-Sitter space (AdS) \cite{Karch:2002sh}.
In the dual picture a falling string corresponds to a cone of energy
propagating along its axis with an opening angle proportional to the
downward angle $\sigma_{0}$ of the string endpoint \cite{Chesler:2014jva,Chesler:2015nqz}.
Such strings can be used as holographic proxies for jets,
if supplied with the required energy and if the string endpoints are
chosen to be moving apart in the center of energy frame. 

We use the model introduced in \cite{Rajagopal:2016uip,Brewer:2017fqy}, which includes input from pQCD to construct an ensemble of initial string conditions that is realistic from a QCD perspective. In this model, the distribution of holographic jet widths is fixed to the pQCD calculation of the variable $C_{1}^{(1)}$ in \cite{Larkoski:2014wba}. A result of \cite{Brewer:2017fqy} is the distribution of energy along strings at late times, which with the width fixes the jet shape completely. This shape matches the measured inclusive jet shape in $pp$ collisions well. 
Note that this model results in null strings, which have the property that the energy distribution within the jet cone stays fixed in vacuum and hence the jet shape in $pp$ does not evolve. For the case with plasma each individual jet widens, but note that this does not necessarily imply that the entire ensemble after applying a transverse momentum cut widens as well \cite{Rajagopal:2016uip}. For null strings quark and gluon jets are the same, except that they have different $C_{1}^{(1)}$ distributions \cite{Larkoski:2014wba}, with gluon jets being wider. We take a 50\% estimate of quark jets for the $p_T$ range studied in this work.
We use the $C_{1}^{(1)}$ distribution to describe the widths of both leading and subleading jets, however we note that the resulting subleading jet shape in vacuum is narrower than measurements \cite{Khachatryan:2016tfj}, presumably due to the presence of third jets. 

We fix the initial vacuum dijet asymmetry as in \cite{Brewer:2017fqy} from a (smeared) \textsc{Pythia}+\textsc{Hydjet} simulation
from \cite{Chatrchyan:2012nia}, which used transverse momentum cuts $(p_{1},\,p_{2})>(120,\,30)$ GeV. 
This initial dijet distribution is realistic in asymmetry, but it is important to note that the system is hence not balanced in transverse momentum. Realistically the subleading jet is often accompanied by a smaller third jet such that the total transverse momentum is balanced. In our study we choose to ignore these third jets, which could have a significant contribution for high $p_T$ particles away from the subleading jet axis.

As in \cite{Brewer:2017fqy} we use a simple blast-wave model for the temperature profile (see \cite{Brewer:2017fqy} for details). Jets are produced according to a binary scaling distribution in the transverse plane and the normalization of the temperature is chosen to lead to reasonable jet suppression. Here we include transverse velocity effects using the fluid-gravity metric. As a simple model for the transverse
fluid velocity we take $u_{i}=-\tanh\left(\frac{4\tau}{3T}\nabla_{i}T\right)$,
which works well for small times \cite{vanderSchee:2012qj,Habich:2014jna}
and gives a realistic magnitude at late times. Here we neglect all gradient terms in the hydrodynamic expansion. It is possible to use a more complete
analysis of string propagation in general metrics, including far-from-equilibrium
dynamics and viscous hydrodynamics, see e.g. \cite{Lekaveckas:2013lha}, but these
corrections do not have significant effects on the results presented here.

\begin{figure*}[t]
\begin{centering}
\includegraphics[width=5.5cm]{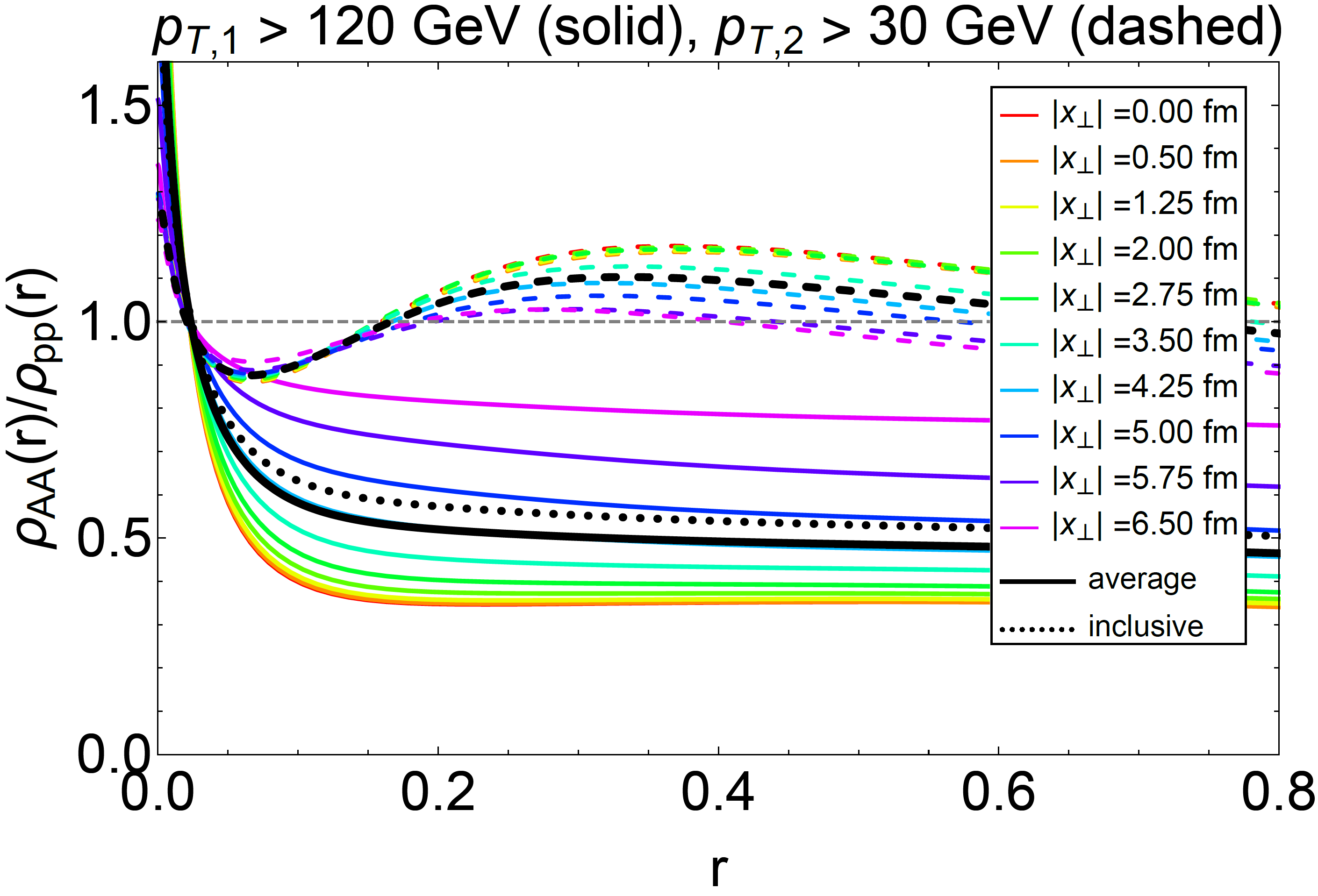}\includegraphics[width=6.25cm]{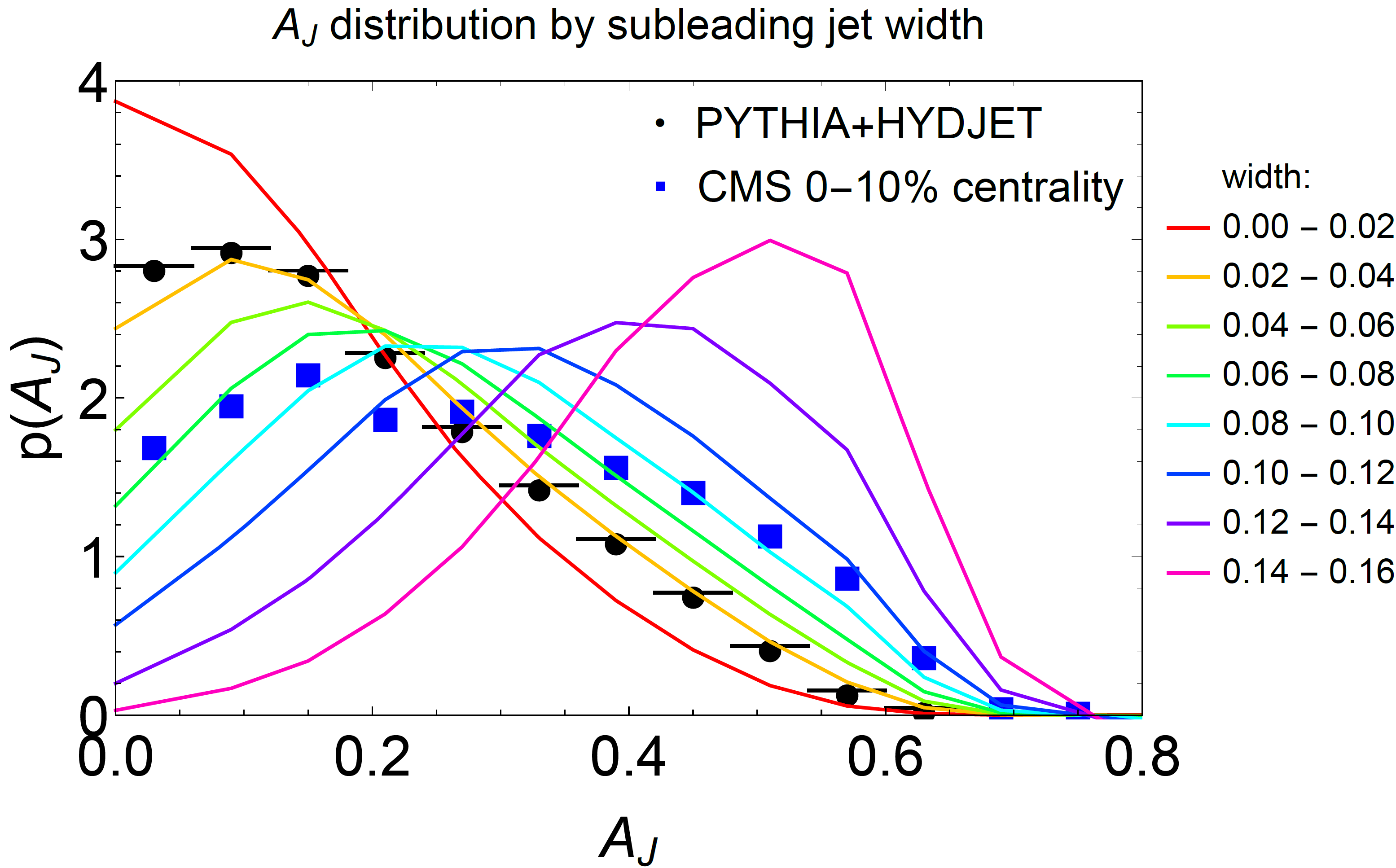}$\,\,$\includegraphics[width=5.25cm]{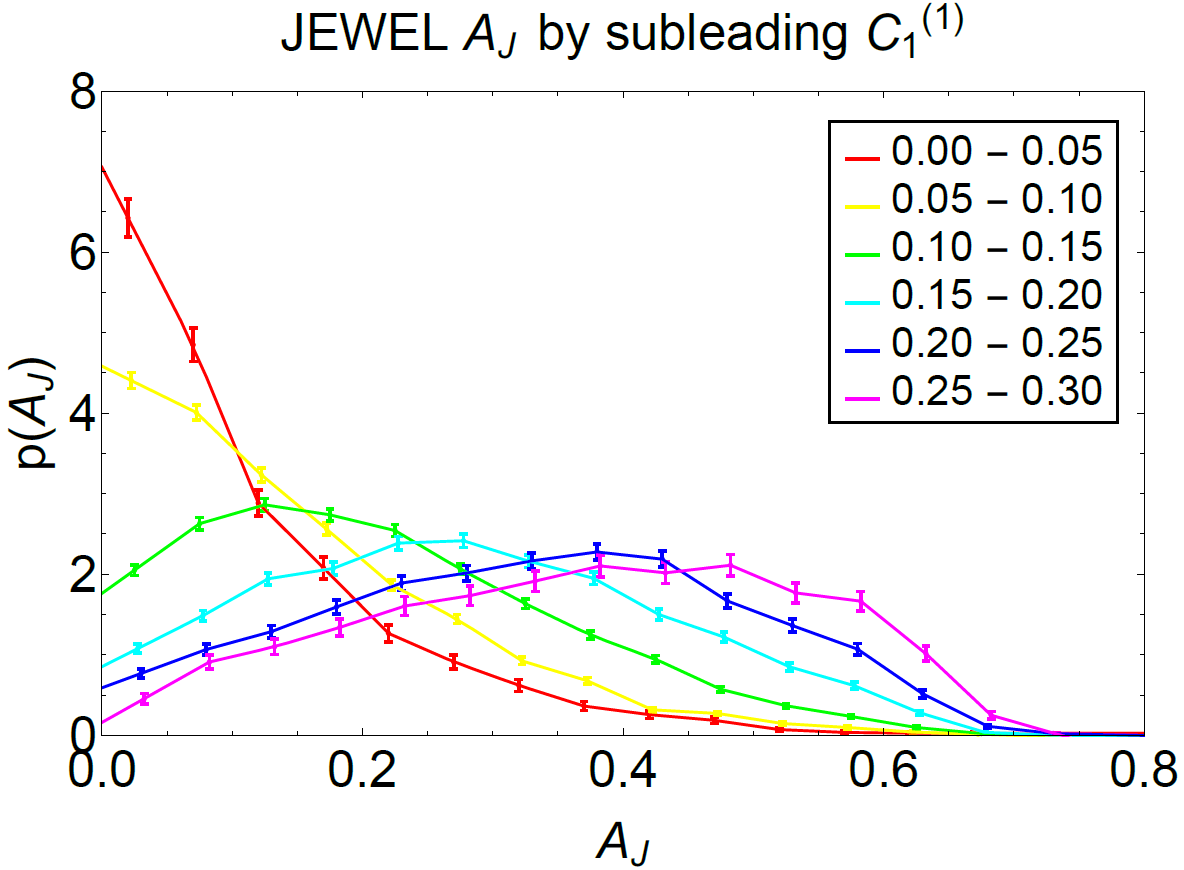}
\par\end{centering}
\caption{(left) Leading (solid) and subleading (dashed) jet shapes are more
heavily modified for centrally produced jets. The average jet shape
modification for leading and subleading jets is shown in black solid
and dashed and for inclusive jets in dotted black. (middle) Our main
result is the dijet asymmetry binned for different widths of the subleading
jets, which shows a strong qualitative difference for wide and narrow
subleading jets (corresponding to unbalanced and balanced dijets,
respectively). The dijets with the most narrow subleading jets are
even more balanced than vacuum dijets. (right) In \textsc{Jewel} dijets with
a wider subleading jet also have a more imbalanced dijet asymmetry
distribution, but for wide subleading jets the distribution is broader
in $A_{J}$ and for narrow subleading jets it is narrower. The errors are statistical only. \label{fig:differentiated}}
\end{figure*}

We start
our strings near the boundary at proper time $\tau_{0}=0.5$ fm/$c$.
For each jet we compute the evolution of a collection of null geodesics, which we translate into an energy
momentum flow when the plasma reaches the freeze-out temperature of
175 MeV \cite{Hatta:2011gh,Chesler:2015nqz}.  
The segments that have
not fallen in the black hole have new AdS angles, which lead to a
new jet shape and new jet $p_{T}$ defined to be the $p_{T}$ within
$r=0.3$ of the average direction of the jet. It is important to note
that strings with endpoints traveling into the bulk at a large downward
angle, or equivalently jets with large width, will fall into the black
hole faster and hence lose more energy to the plasma than their narrow
counterparts.

\noindent \textbf{Dijets and jet shapes} - The energy loss of dijets
in HIC has particularly rich and interesting physics. Depending on
the production point of the dijet the two jets traverse a different
part of the medium. Both jets may also lose energy differently, depending
on their width in the holographic picture or on the particular structure
of the particle shower in a weakly coupled picture.

Fig. \ref{fig:RAA} (left) shows the resulting suppression of jets
of a given $p_{T}$ relative to the initial ensemble (the jet nuclear
modification factor $R_{AA}^{{\rm jet}}$), for all jets in the ensemble
separated by their initial positions. Jets produced at the center
lose much more energy on average than jets produced at the edge and
hence are more suppressed. In order to quantify the effects of the
path length imbalance and the jet width fluctuations on the $A_{J}$
distribution we show the distribution for the ensemble separately
for different initial positions (dijets originating at the center
always have balanced path lengths), see Fig. \ref{fig:RAA} (middle),
and subsequently for the subset of these dijets where the two jets
have the same jet width, see Fig. \ref{fig:RAA} (right). In this
case the jets in a dijet system do not have independent energy loss
fluctuations. When the dijets have both widths varying independently
the asymmetry does not depend strongly on the path length, confirming
results of \cite{Milhano:2015mng}. However, for the theoretical
case that both jets in a dijet have the same width the path length
imbalance is crucial to obtain the dijet imbalance. It is surprising
to note that on average the final $A_{J}$ distribution is almost
identical regardless of whether the jet widths fluctuate independently
or not, and furthermore that the fluctuating dijets created at the edge ($x=6.5\,$ fm) are in fact less asymmetric than the fluctuating ones. The dijet asymmetry is also almost identical to the one obtained for centrally-produced
dijets. 
%
This surprising finding in particular implies that the dijet distribution is rather robust and that describing energy loss fluctuations is not crucial for obtaining the correct dijet asymmetry distribution. However, this is not the case for more sophisticated observables discussed below.

To study both effects more closely we show the jet shape modifications
for both subleading and leading jets in Fig. \ref{fig:differentiated}
(left), where $\rho_{AA,pp}(r)$ are the normalized transverse momentum
distributions as a function of the angular distance $r$ to the average
direction of the jet for $AA$ and $pp$ collisions respectively.
Clearly, jets produced at the center also have their shapes more strongly
modified, which in the case of independent energy loss makes them
more imbalanced. This gives an extra explanation why the path length
is not crucial for the dijet asymmetry distribution \cite{Milhano:2015mng}:
jets that have the same path length often are jets that are modified
most which makes them more unbalanced for that reason.

\begin{figure*}[t]
\begin{centering}
\includegraphics[width=6cm]{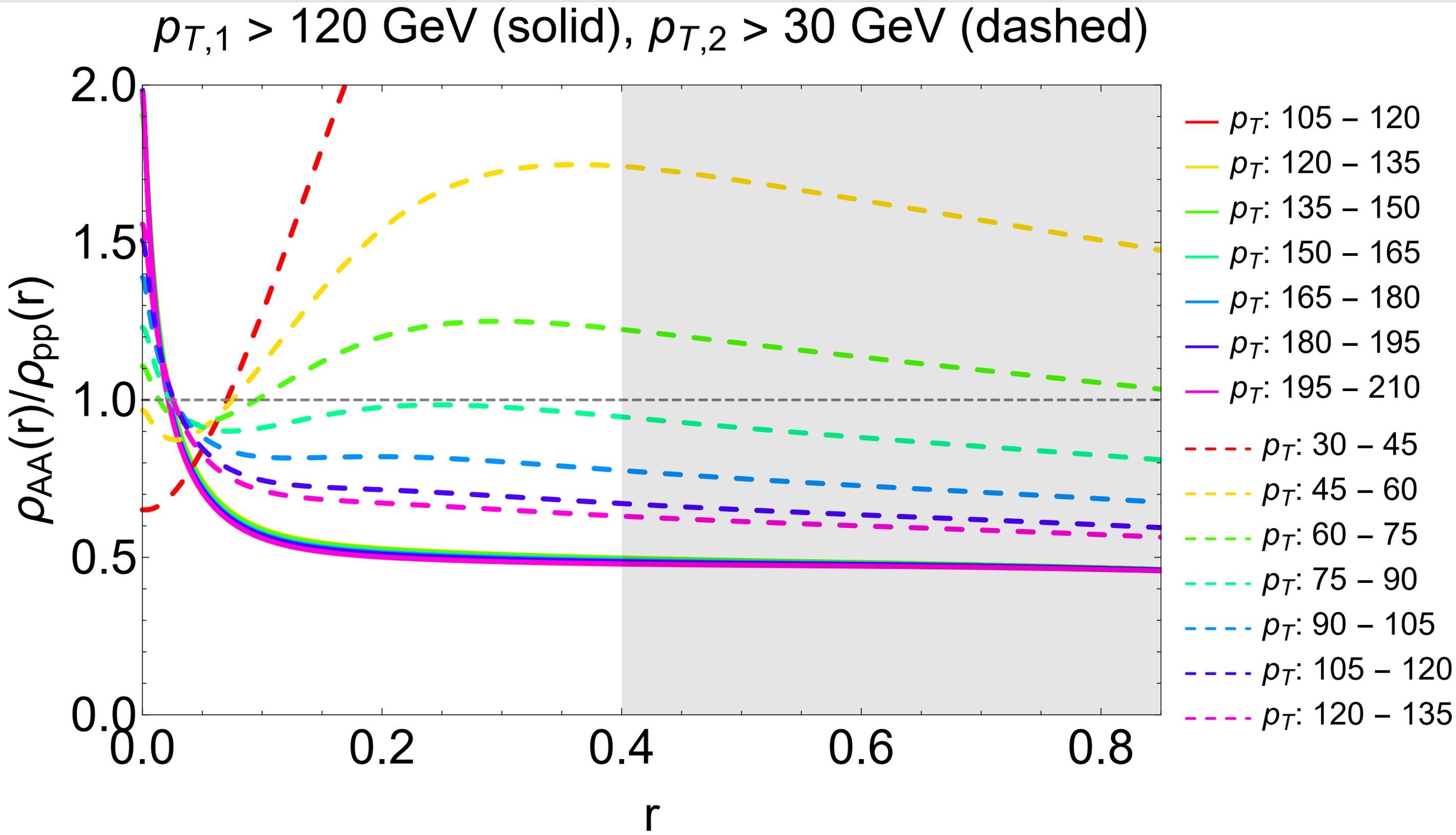}\includegraphics[width=6cm]{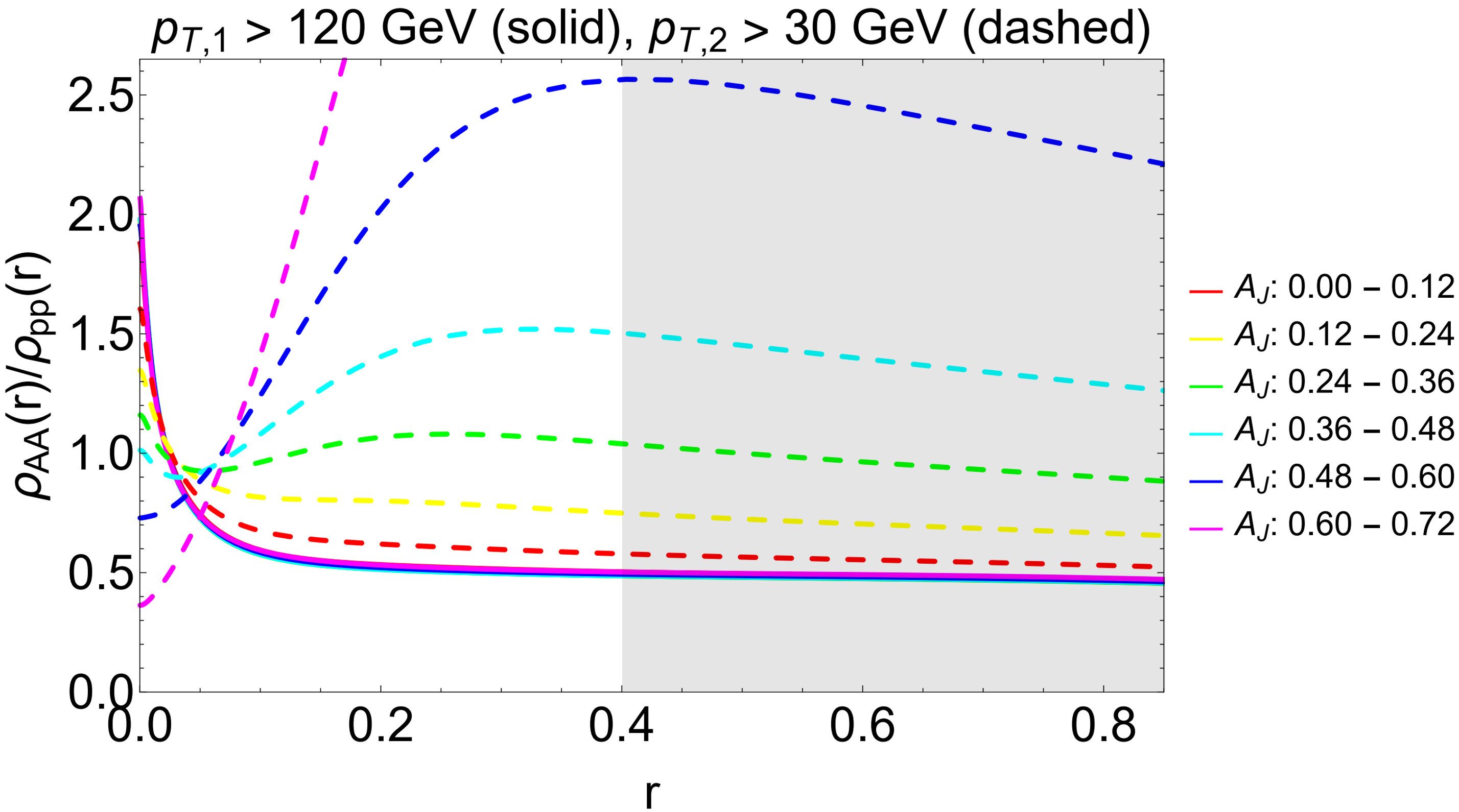}\includegraphics[width=6cm]{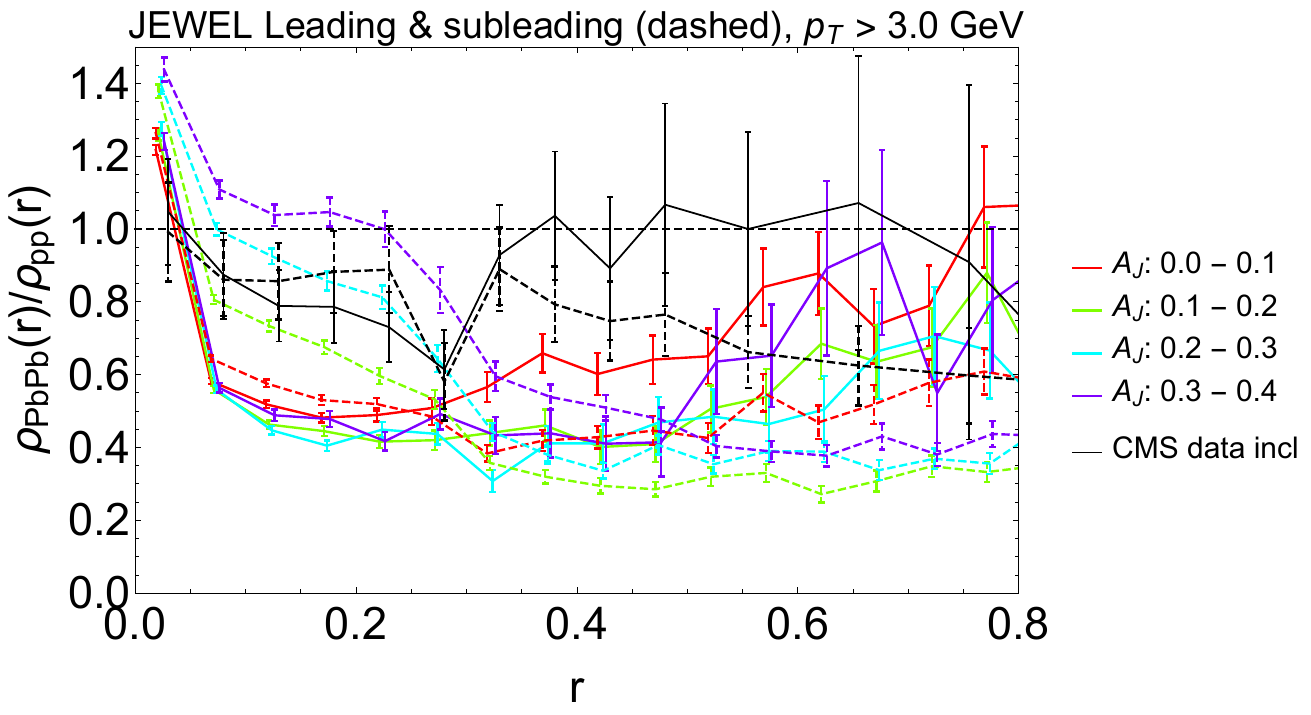}
\par\end{centering}
\caption{We show the leading (solid) and subleading (dashed) jet shape modifications
binned for different jet transverse momentum (left) and different
dijet imbalance $A_{J}$ (middle) for the holographic model. (right)
shows the equivalent plot in \textsc{Jewel}, overlayed with CMS data  
obtained from \textcolor{black}{\cite{Khachatryan:2016tfj}}, both considering only particles with $p_{T}>3$ GeV. The right
figure uses $p_{2}>50$ GeV instead of $p_{2}>30$ GeV to be consistent
with \textcolor{black}{\cite{Khachatryan:2016tfj}}, which leads
to small differences for $A_{J}>0.41$. It is clear that the leading
jet shapes consistently get narrower. The shape of the subleading
jet depends on both $A_{J}$ and $p_{T}$, with less balanced events
or smaller subleading jet $p_{T}$ giving wider subleading jets.\label{fig:jet-shapes}}
\end{figure*}

Our main result is to select jets on their final width (defined here
as $w\equiv\int_{0}^{R=0.3}r\rho(r) {\rm d}r$), which unlike selecting on
starting positions is also possible experimentally. Fig. \ref{fig:differentiated}
(middle) shows the $A_{J}$ distribution binned for different widths
of the subleading jet. Clearly, the width of the subleading jet is
essential in the dijet asymmetry: wider subleading jets lead to asymmetric
dijets, whereas the narrowest subleading jets even lead to a distribution
that is more symmetric than the original $pp$ asymmetry. In Fig.
\ref{fig:differentiated} (right) we compare our result with results
from the \textsc{Jewel} Monte Carlo generator, which is based on a weakly-coupled
kinetic theory with scattering centers in the QGP \cite{Zapp:2008gi}.
We generate events at $2.76$ TeV with the default medium model chosen
to match hydrodynamic simulations at this energy and reconstruct anti-$k_{t}$
jets using FastJet. We turn off medium recoils in \textsc{Jewel} so that medium
back reaction is ignored as in the holographic calculation. Wider
subleading jets are also more imbalanced on average in \textsc{Jewel}, but
for larger $C_{1}^{(1)}$ values the distribution is wider than in
the holographic case.

The jet shapes in the leftmost panel of Fig. \ref{fig:differentiated} are `theoretical',
in the sense that experimentally it is impossible to assign a starting
location $x_{1}$ to an individual jet. It is however possible to
see a similar effect experimentally by selecting dijets according
to their transverse momentum (Fig. \ref{fig:jet-shapes}, left) or
their asymmetry (Fig. \ref{fig:jet-shapes}, middle). Quite curiously,
the leading jet shape does not depend strongly on either $p_{T}$
or $A_{J}$ (an experimental indication that this is correct can be
found in \cite{Khachatryan:2016tfj}, see also our discussion section below). However, we find
that the subleading jet shapes strongly depend on both the transverse
momentum and the dijet asymmetry, with more unbalanced dijets having
wider subleading jets. The narrowing of leading jets and widening of the subleading jets is in agreement
with the two competing effects studied in \cite{Rajagopal:2016uip},
where every jet gets wider, but the average shape of the ensemble
of jets can narrow due to the steeply falling jet production spectrum.
For the leading jets the second effect is dominant, but for subleading jets the first effect dominates, since the subleading
$p_{T}$ cut is much lower than the typical $p_{T}$ of a subleading
jet. Subleading jets therefore usually stay within the sample and especially for larger dijet asymmetries gain considerably in width. 
For qualitative illustration, we show in grey in Fig. \ref{fig:jet-shapes} regions where our results are only partial, since in those regions both medium backreaction (for low $p_T$ hadrons) and third jets for the subleading jets (high $p_T$ hadrons) can significantly modify the jet shapes.  

The \textsc{Jewel} analysis (Fig. \ref{fig:jet-shapes}, right) confirms the
jet shape dependence on $A_{J}$ qualitatively, except that at large
$r$ the subleading jet shapes in \textsc{Jewel} are narrower. One advantage
of \textsc{Jewel} is that it contains a full Monte Carlo, and hence incorporates
3rd jets, or incoherent partons at relatively large $r$ that lose
energy to QGP independently (see i.e. \cite{MehtarTani:2011tz,Hulcher:2017cpt}).
These type of partons are a rather typical contribution to subleading
jet shapes at large $r$, and the quenching of such partons will lead
to narrower subleading jet shapes. Since this is not taken into account
in the holographic model the subleading jet shape at large $r$ cannot
be accurately compared with experimental data. The \textsc{Jewel} analysis,
on the other hand, compares quite well with CMS jet shape data \cite{Khachatryan:2016tfj},
when restricting to $p_{T}>3$ GeV (this leaves out the thermal particles
that we did not take into account). This effect is likely also related
to the reason that \textsc{Jewel} has wider distributions in Fig. \ref{fig:differentiated}
(right), since incoherent partons do not give rise to subleading jets
as wide as in the holographic model. However, the width of the jet
is less sensitive to the large $r$ part of the jet shape, so that
the middle of Fig. \ref{fig:differentiated} is more robust than the
subleading jet shape at large $r$.

Qualitatively the dependence on the subleading jet width is intuitive:
a wider jet generically loses more energy than a narrow jet, and hence
will likely end up with significantly less energy than the other jet.
The full story is however more complicated: the combination of the
steeply falling jet spectrum and momentum cuts biases the interpretation
of jet modification observables and necessitates considering an ensemble
of jets \cite{Rajagopal:2016uip,Brewer:2017fqy}. Furthermore in
both \textsc{Jewel} and holography the modified jet width can deviate substantially
from the initial jet width. Nevertheless,
both the holographic model and \textsc{Jewel} agree with the naive intuition,
although the holographic model produces significantly wider subleading
jets. This suggests either a stronger influence of the jet width on
the energy loss in holography, or a stronger jet width evolution while
the jet traverses the plasma.

\noindent \textbf{Discussion} - In this work we take input from pQCD
to model dijet evolution through a holographic plasma. This suggests
two interesting observables requiring further study: the $A_{J}$
distribution binned for different subleading jet widths (Fig. \ref{fig:differentiated},
middle) and the jet shape modifications of leading and subleading
jets binned for different $A_{J}$ (Fig. \ref{fig:jet-shapes}, middle).
The former showed similar qualitative features to the results from
\textsc{Jewel} Monte Carlo, while the jet shapes agreed qualitatively for the
leading jet shapes and for small $r$. Both experimental measurements
and different model results of these observables (including pQCD computations,
for example \cite{He:2011pd}) will shed light on the interplay of
path lengths, jet energy loss fluctuations, jet structure and substructure.

The model presented requires many future improvements, and for that
reason the results presented should be seen as qualitative. We only
consider back-to-back dijets without considering third jets
\footnote{Note that our dijets follow the measured dijet asymmetry distribution and that transverse momentum is hence not balanced.}
, and we do not treat the back reaction of jets on the medium or include any
hadronization procedure. All of these can potentially be improved
by incorporating the model in a Monte Carlo framework such as \textsc{Jetscape}
\cite{Cao:2017zih} (see also \cite{Casalderrey-Solana:2016jvj}).
We stress in particular that third jets can affect the subleading
jet shapes at intermediate to large $r$, which likely contributes
to the difference between our result (Fig. \ref{fig:jet-shapes} middle)
and \textsc{Jewel} (Fig. \ref{fig:jet-shapes} right). We note that the averaged
subleading jet shape of \textsc{Jewel} agreed quite well with the result of
CMS when including particles with $p_{T}>3$ GeV. This cut in transverse
momentum is important to separate the jet from the medium (also done
in \cite{Sirunyan:2018jqr}). However it is known that recoils have a large effect on jet shapes when lower $p_T$ particles are included \cite{KunnawalkamElayavalli:2017hxo}, and they may have some impact on the jet shapes for $p_T>3$ GeV especially at large $r$. Finally, it would also be interesting
to extend this analysis to $\gamma-$jet events, where it is possible
to use the photon as a probe of the initial jet energy and directly
study how the shape modification of a jet depends on its energy loss.

\begin{figure}[t]
\begin{centering}
\includegraphics[width=9cm]{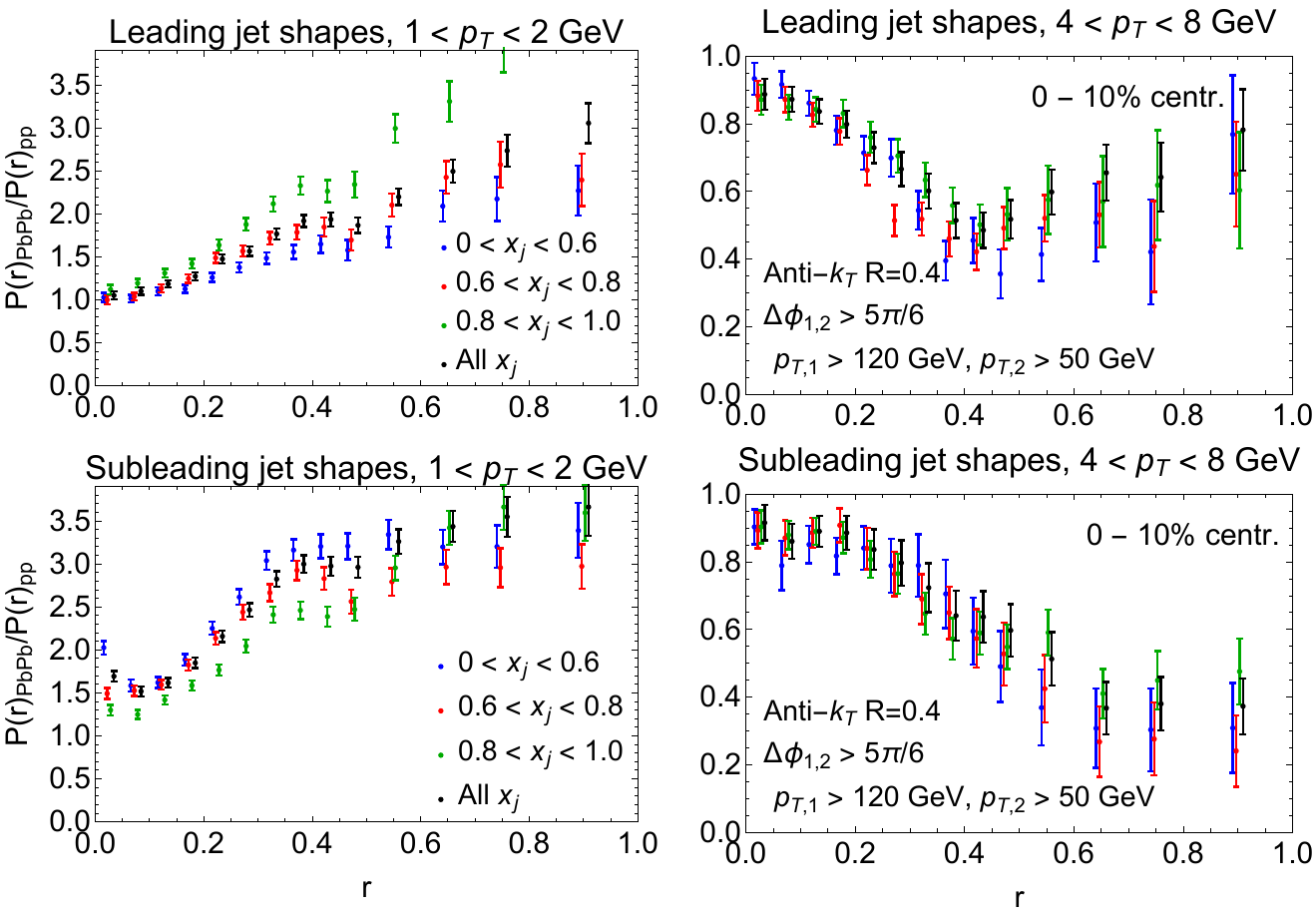}
\par\end{centering}
\caption{We show the leading (top) and subleading (bottom) jet shape modifications derived from CMS data \cite{Sirunyan:2021jty}
binned for different dijet asymmetries $x_j = p_{T,2} / p_{T,1}$ for charged hadrons with $p_T$ within $(1,\,2)\,$GeV (left) and $(4,\,8)\,$GeV (right).
Unbalanced jets (blue) lead to more low $p_T$ particles for subleading jets, whereas both leading and subleading jets have a reduced number of high $p_T$ hadrons at large angles.\label{fig:CMSdata}}
\end{figure}


Even though as argued our results have to be taken qualitatively, it is still interesting to compare our model predictions with the recent measurements by CMS \cite{Sirunyan:2021jty}.
Though a preprint of this work appeared before the data was taken, for completeness we show in Fig.~\ref{fig:CMSdata} ratios derived from the recent measurements by CMS \cite{Sirunyan:2021jty} of the leading and subleading jet shapes of low and high $p_T$ particles as a function of the dijet asymmetry. 
Note that this CMS measurement measures the absolute shapes $P(r)$ instead of the self-normalised shapes $\rho(r)$ as in Fig.~\ref{fig:jet-shapes}, which in particular means that the $P(r)$ ratios do not necessarily cross unity.
Error bars shown are the quadrature sum of statistical and systematic uncertainties, which in this case are dominated by systematics.
These $p_T$-binned ratios can give more insights than the $p_T$-integrated results of \cite{Sirunyan:2021jty}, since the different physical scales can separate medium backreaction and the quenching of hard partons (including third jets).

Optimistically, the high $4<p_T<8$ bin behaves qualitatively similarly to our leading jet shape predictions and those from \textsc{Jewel} shown in Fig.~\ref{fig:jet-shapes}, whereas the subleading jet shapes at $1<p_T<2$~GeV also qualitatively follow the trend predicted in holography. Indeed, at high $p_T$ the leading jet shapes become narrower and depend only weakly on the dijet asymmetry, while the subleading jet shapes at low $p_T$ are ordered in the same way as in Fig. \ref{fig:jet-shapes} -- more unbalanced dijets (smaller $x_j$) correspond to wider subleading jets. 
It is reasonable that the subleading jet broadening in Fig.~\ref{fig:jet-shapes} shows up at low $p_T$, but we note that this is also the regime where we can expect effects from medium response that we do not include here.
Contrary to our expectations from Fig.~\ref{fig:jet-shapes}, it is interesting that the subleading jet shapes at high $p_T$ do not depend significantly on the asymmetry and become narrower. At large $r$, the subleading jet shapes in \textsc{Jewel} also become narrower. We expect that third jets may be crucial for understanding the lack of dependence of the subleading jet shape on the asymmetry for high-$p_T$ particles.


At the moment both the holographic model (which is more strongly coupled) and \textsc{Jewel} (which is based on weak coupling)  are too simple to fully describe the experimental data. The fact that for these two theories this observable is much more difficult to obtain  than the dijet asymmetry distribution by itself 
(see Fig.~\ref{fig:RAA}) makes it of special interest for future work, since it could be used to differentiate between the regimes.


\begin{acknowledgments}
\noindent \textbf{Acknowledgments} - We thank Olga Evdokimov, Yen-Jie Lee, Krishna Rajagopal, Marta
Verweij and Ivan Vitev for interesting discussions. WS is supported
by VENI grant 680-47-458 from the Netherlands Organisation for Scientific
Research (NWO). WS and JB are supported by the U.S. Department of
Energy under grant Contract Number DE-SC0011090. AS is partially supported
through the LANL/LDRD Program. AS is also grateful for support
from RFBR grant 17-02-01108 and RFBR  grant  18-02-40056.
\end{acknowledgments}

\bibliographystyle{apsrev4-1}
\bibliography{prl11revised}

\begin{thebibliography}{37}%
\makeatletter
\providecommand \@ifxundefined [1]{%
 \@ifx{#1\undefined}
}%
\providecommand \@ifnum [1]{%
 \ifnum #1\expandafter \@firstoftwo
 \else \expandafter \@secondoftwo
 \fi
}%
\providecommand \@ifx [1]{%
 \ifx #1\expandafter \@firstoftwo
 \else \expandafter \@secondoftwo
 \fi
}%
\providecommand \natexlab [1]{#1}%
\providecommand \enquote  [1]{``#1''}%
\providecommand \bibnamefont  [1]{#1}%
\providecommand \bibfnamefont [1]{#1}%
\providecommand \citenamefont [1]{#1}%
\providecommand \href@noop [0]{\@secondoftwo}%
\providecommand \href [0]{\begingroup \@sanitize@url \@href}%
\providecommand \@href[1]{\@@startlink{#1}\@@href}%
\providecommand \@@href[1]{\endgroup#1\@@endlink}%
\providecommand \@sanitize@url [0]{\catcode `\\12\catcode `\$12\catcode
  `\&12\catcode `\#12\catcode `\^12\catcode `\_12\catcode `\%12\relax}%
\providecommand \@@startlink[1]{}%
\providecommand \@@endlink[0]{}%
\providecommand \url  [0]{\begingroup\@sanitize@url \@url }%
\providecommand \@url [1]{\endgroup\@href {#1}{\urlprefix }}%
\providecommand \urlprefix  [0]{URL }%
\providecommand \Eprint [0]{\href }%
\providecommand \doibase [0]{http://dx.doi.org/}%
\providecommand \selectlanguage [0]{\@gobble}%
\providecommand \bibinfo  [0]{\@secondoftwo}%
\providecommand \bibfield  [0]{\@secondoftwo}%
\providecommand \translation [1]{[#1]}%
\providecommand \BibitemOpen [0]{}%
\providecommand \bibitemStop [0]{}%
\providecommand \bibitemNoStop [0]{.\EOS\space}%
\providecommand \EOS [0]{\spacefactor3000\relax}%
\providecommand \BibitemShut  [1]{\csname bibitem#1\endcsname}%
\let\auto@bib@innerbib\@empty
\bibitem [{\citenamefont {Casalderrey-Solana}\ \emph
  {et~al.}(2014)\citenamefont {Casalderrey-Solana}, \citenamefont {Liu},
  \citenamefont {Mateos}, \citenamefont {Rajagopal},\ and\ \citenamefont
  {Wiedemann}}]{CasalderreySolana:2011us}%
  \BibitemOpen
  \bibfield  {author} {\bibinfo {author} {\bibfnamefont {J.}~\bibnamefont
  {Casalderrey-Solana}}, \bibinfo {author} {\bibfnamefont {H.}~\bibnamefont
  {Liu}}, \bibinfo {author} {\bibfnamefont {D.}~\bibnamefont {Mateos}},
  \bibinfo {author} {\bibfnamefont {K.}~\bibnamefont {Rajagopal}}, \ and\
  \bibinfo {author} {\bibfnamefont {U.~A.}\ \bibnamefont {Wiedemann}},\ }\href
  {\doibase 10.1017/CBO9781139136747} {\emph {\bibinfo {title} {{Gauge/String
  Duality, Hot QCD and Heavy Ion Collisions}}}}\ (\bibinfo  {publisher}
  {Cambridge University Press},\ \bibinfo {year} {2014})\ \Eprint
  {http://arxiv.org/abs/1101.0618} {arXiv:1101.0618 [hep-th]} \BibitemShut
  {NoStop}%
\bibitem [{\citenamefont {Busza}\ \emph {et~al.}(2018)\citenamefont {Busza},
  \citenamefont {Rajagopal},\ and\ \citenamefont {van~der
  Schee}}]{Busza:2018rrf}%
  \BibitemOpen
  \bibfield  {author} {\bibinfo {author} {\bibfnamefont {W.}~\bibnamefont
  {Busza}}, \bibinfo {author} {\bibfnamefont {K.}~\bibnamefont {Rajagopal}}, \
  and\ \bibinfo {author} {\bibfnamefont {W.}~\bibnamefont {van~der Schee}},\
  }\href {\doibase 10.1146/annurev-nucl-101917-020852} {\bibfield  {journal}
  {\bibinfo  {journal} {Ann. Rev. Nucl. Part. Sci.}\ }\textbf {\bibinfo
  {volume} {68}},\ \bibinfo {pages} {339} (\bibinfo {year} {2018})},\ \Eprint
  {http://arxiv.org/abs/1802.04801} {arXiv:1802.04801 [hep-ph]} \BibitemShut
  {NoStop}%
\bibitem [{\citenamefont {Connors}\ \emph {et~al.}(2018)\citenamefont
  {Connors}, \citenamefont {Nattrass}, \citenamefont {Reed},\ and\
  \citenamefont {Salur}}]{Connors:2017ptx}%
  \BibitemOpen
  \bibfield  {author} {\bibinfo {author} {\bibfnamefont {M.}~\bibnamefont
  {Connors}}, \bibinfo {author} {\bibfnamefont {C.}~\bibnamefont {Nattrass}},
  \bibinfo {author} {\bibfnamefont {R.}~\bibnamefont {Reed}}, \ and\ \bibinfo
  {author} {\bibfnamefont {S.}~\bibnamefont {Salur}},\ }\href {\doibase
  10.1103/RevModPhys.90.025005} {\bibfield  {journal} {\bibinfo  {journal}
  {Rev. Mod. Phys.}\ }\textbf {\bibinfo {volume} {90}},\ \bibinfo {pages}
  {025005} (\bibinfo {year} {2018})},\ \Eprint
  {http://arxiv.org/abs/1705.01974} {arXiv:1705.01974 [nucl-ex]} \BibitemShut
  {NoStop}%
\bibitem [{\citenamefont {Rajagopal}\ \emph {et~al.}(2016)\citenamefont
  {Rajagopal}, \citenamefont {Sadofyev},\ and\ \citenamefont {van~der
  Schee}}]{Rajagopal:2016uip}%
  \BibitemOpen
  \bibfield  {author} {\bibinfo {author} {\bibfnamefont {K.}~\bibnamefont
  {Rajagopal}}, \bibinfo {author} {\bibfnamefont {A.~V.}\ \bibnamefont
  {Sadofyev}}, \ and\ \bibinfo {author} {\bibfnamefont {W.}~\bibnamefont
  {van~der Schee}},\ }\href {\doibase 10.1103/PhysRevLett.116.211603}
  {\bibfield  {journal} {\bibinfo  {journal} {Phys. Rev. Lett.}\ }\textbf
  {\bibinfo {volume} {116}},\ \bibinfo {pages} {211603} (\bibinfo {year}
  {2016})},\ \Eprint {http://arxiv.org/abs/1602.04187} {arXiv:1602.04187
  [nucl-th]} \BibitemShut {NoStop}%
\bibitem [{\citenamefont {Brewer}\ \emph {et~al.}(2018)\citenamefont {Brewer},
  \citenamefont {Rajagopal}, \citenamefont {Sadofyev},\ and\ \citenamefont {Van
  Der~Schee}}]{Brewer:2017fqy}%
  \BibitemOpen
  \bibfield  {author} {\bibinfo {author} {\bibfnamefont {J.}~\bibnamefont
  {Brewer}}, \bibinfo {author} {\bibfnamefont {K.}~\bibnamefont {Rajagopal}},
  \bibinfo {author} {\bibfnamefont {A.}~\bibnamefont {Sadofyev}}, \ and\
  \bibinfo {author} {\bibfnamefont {W.}~\bibnamefont {Van Der~Schee}},\ }\href
  {\doibase 10.1007/JHEP02(2018)015} {\bibfield  {journal} {\bibinfo  {journal}
  {JHEP}\ }\textbf {\bibinfo {volume} {02}},\ \bibinfo {pages} {015} (\bibinfo
  {year} {2018})},\ \Eprint {http://arxiv.org/abs/1710.03237} {arXiv:1710.03237
  [nucl-th]} \BibitemShut {NoStop}%
\bibitem [{\citenamefont {Milhano}\ and\ \citenamefont
  {Zapp}(2016)}]{Milhano:2015mng}%
  \BibitemOpen
  \bibfield  {author} {\bibinfo {author} {\bibfnamefont {J.~G.}\ \bibnamefont
  {Milhano}}\ and\ \bibinfo {author} {\bibfnamefont {K.~C.}\ \bibnamefont
  {Zapp}},\ }\href {\doibase 10.1140/epjc/s10052-016-4130-9} {\bibfield
  {journal} {\bibinfo  {journal} {Eur. Phys. J. C}\ }\textbf {\bibinfo {volume}
  {76}},\ \bibinfo {pages} {288} (\bibinfo {year} {2016})},\ \Eprint
  {http://arxiv.org/abs/1512.08107} {arXiv:1512.08107 [hep-ph]} \BibitemShut
  {NoStop}%
\bibitem [{\citenamefont {Escobedo}\ and\ \citenamefont
  {Iancu}(2016)}]{Escobedo:2016jbm}%
  \BibitemOpen
  \bibfield  {author} {\bibinfo {author} {\bibfnamefont {M.~A.}\ \bibnamefont
  {Escobedo}}\ and\ \bibinfo {author} {\bibfnamefont {E.}~\bibnamefont
  {Iancu}},\ }\href {\doibase 10.1007/JHEP05(2016)008} {\bibfield  {journal}
  {\bibinfo  {journal} {JHEP}\ }\textbf {\bibinfo {volume} {05}},\ \bibinfo
  {pages} {008} (\bibinfo {year} {2016})},\ \Eprint
  {http://arxiv.org/abs/1601.03629} {arXiv:1601.03629 [hep-ph]} \BibitemShut
  {NoStop}%
\bibitem [{\citenamefont {Chesler}\ and\ \citenamefont
  {Rajagopal}(2016)}]{Chesler:2015nqz}%
  \BibitemOpen
  \bibfield  {author} {\bibinfo {author} {\bibfnamefont {P.~M.}\ \bibnamefont
  {Chesler}}\ and\ \bibinfo {author} {\bibfnamefont {K.}~\bibnamefont
  {Rajagopal}},\ }\href {\doibase 10.1007/JHEP05(2016)098} {\bibfield
  {journal} {\bibinfo  {journal} {JHEP}\ }\textbf {\bibinfo {volume} {05}},\
  \bibinfo {pages} {098} (\bibinfo {year} {2016})},\ \Eprint
  {http://arxiv.org/abs/1511.07567} {arXiv:1511.07567 [hep-th]} \BibitemShut
  {NoStop}%
\bibitem [{\citenamefont {Chien}\ and\ \citenamefont
  {Vitev}(2016)}]{Chien:2015hda}%
  \BibitemOpen
  \bibfield  {author} {\bibinfo {author} {\bibfnamefont {Y.-T.}\ \bibnamefont
  {Chien}}\ and\ \bibinfo {author} {\bibfnamefont {I.}~\bibnamefont {Vitev}},\
  }\href {\doibase 10.1007/JHEP05(2016)023} {\bibfield  {journal} {\bibinfo
  {journal} {JHEP}\ }\textbf {\bibinfo {volume} {05}},\ \bibinfo {pages} {023}
  (\bibinfo {year} {2016})},\ \Eprint {http://arxiv.org/abs/1509.07257}
  {arXiv:1509.07257 [hep-ph]} \BibitemShut {NoStop}%
\bibitem [{\citenamefont {Casalderrey-Solana}\ \emph
  {et~al.}(2017)\citenamefont {Casalderrey-Solana}, \citenamefont {Gulhan},
  \citenamefont {Milhano}, \citenamefont {Pablos},\ and\ \citenamefont
  {Rajagopal}}]{Casalderrey-Solana:2016jvj}%
  \BibitemOpen
  \bibfield  {author} {\bibinfo {author} {\bibfnamefont {J.}~\bibnamefont
  {Casalderrey-Solana}}, \bibinfo {author} {\bibfnamefont {D.}~\bibnamefont
  {Gulhan}}, \bibinfo {author} {\bibfnamefont {G.}~\bibnamefont {Milhano}},
  \bibinfo {author} {\bibfnamefont {D.}~\bibnamefont {Pablos}}, \ and\ \bibinfo
  {author} {\bibfnamefont {K.}~\bibnamefont {Rajagopal}},\ }\href {\doibase
  10.1007/JHEP03(2017)135} {\bibfield  {journal} {\bibinfo  {journal} {JHEP}\
  }\textbf {\bibinfo {volume} {03}},\ \bibinfo {pages} {135} (\bibinfo {year}
  {2017})},\ \Eprint {http://arxiv.org/abs/1609.05842} {arXiv:1609.05842
  [hep-ph]} \BibitemShut {NoStop}%
\bibitem [{\citenamefont {Kunnawalkam~Elayavalli}\ and\ \citenamefont
  {Zapp}(2017)}]{KunnawalkamElayavalli:2017hxo}%
  \BibitemOpen
  \bibfield  {author} {\bibinfo {author} {\bibfnamefont {R.}~\bibnamefont
  {Kunnawalkam~Elayavalli}}\ and\ \bibinfo {author} {\bibfnamefont {K.~C.}\
  \bibnamefont {Zapp}},\ }\href {\doibase 10.1007/JHEP07(2017)141} {\bibfield
  {journal} {\bibinfo  {journal} {JHEP}\ }\textbf {\bibinfo {volume} {07}},\
  \bibinfo {pages} {141} (\bibinfo {year} {2017})},\ \Eprint
  {http://arxiv.org/abs/1707.01539} {arXiv:1707.01539 [hep-ph]} \BibitemShut
  {NoStop}%
\bibitem [{\citenamefont {Luo}\ \emph {et~al.}(2018)\citenamefont {Luo},
  \citenamefont {Cao}, \citenamefont {He},\ and\ \citenamefont
  {Wang}}]{Luo:2018pto}%
  \BibitemOpen
  \bibfield  {author} {\bibinfo {author} {\bibfnamefont {T.}~\bibnamefont
  {Luo}}, \bibinfo {author} {\bibfnamefont {S.}~\bibnamefont {Cao}}, \bibinfo
  {author} {\bibfnamefont {Y.}~\bibnamefont {He}}, \ and\ \bibinfo {author}
  {\bibfnamefont {X.-N.}\ \bibnamefont {Wang}},\ }\href {\doibase
  10.1016/j.physletb.2018.06.025} {\bibfield  {journal} {\bibinfo  {journal}
  {Phys. Lett. B}\ }\textbf {\bibinfo {volume} {782}},\ \bibinfo {pages} {707}
  (\bibinfo {year} {2018})},\ \Eprint {http://arxiv.org/abs/1803.06785}
  {arXiv:1803.06785 [hep-ph]} \BibitemShut {NoStop}%
\bibitem [{\citenamefont {He}\ \emph {et~al.}(2019)\citenamefont {He},
  \citenamefont {Cao}, \citenamefont {Chen}, \citenamefont {Luo}, \citenamefont
  {Pang},\ and\ \citenamefont {Wang}}]{He:2018xjv}%
  \BibitemOpen
  \bibfield  {author} {\bibinfo {author} {\bibfnamefont {Y.}~\bibnamefont
  {He}}, \bibinfo {author} {\bibfnamefont {S.}~\bibnamefont {Cao}}, \bibinfo
  {author} {\bibfnamefont {W.}~\bibnamefont {Chen}}, \bibinfo {author}
  {\bibfnamefont {T.}~\bibnamefont {Luo}}, \bibinfo {author} {\bibfnamefont
  {L.-G.}\ \bibnamefont {Pang}}, \ and\ \bibinfo {author} {\bibfnamefont
  {X.-N.}\ \bibnamefont {Wang}},\ }\href {\doibase 10.1103/PhysRevC.99.054911}
  {\bibfield  {journal} {\bibinfo  {journal} {Phys. Rev. C}\ }\textbf {\bibinfo
  {volume} {99}},\ \bibinfo {pages} {054911} (\bibinfo {year} {2019})},\
  \Eprint {http://arxiv.org/abs/1809.02525} {arXiv:1809.02525 [nucl-th]}
  \BibitemShut {NoStop}%
\bibitem [{\citenamefont {Pablos}(2020)}]{Pablos:2019ngg}%
  \BibitemOpen
  \bibfield  {author} {\bibinfo {author} {\bibfnamefont {D.}~\bibnamefont
  {Pablos}},\ }\href {\doibase 10.1103/PhysRevLett.124.052301} {\bibfield
  {journal} {\bibinfo  {journal} {Phys. Rev. Lett.}\ }\textbf {\bibinfo
  {volume} {124}},\ \bibinfo {pages} {052301} (\bibinfo {year} {2020})},\
  \Eprint {http://arxiv.org/abs/1907.12301} {arXiv:1907.12301 [hep-ph]}
  \BibitemShut {NoStop}%
\bibitem [{\citenamefont {Casalderrey-Solana}\ \emph
  {et~al.}(2020)\citenamefont {Casalderrey-Solana}, \citenamefont {Milhano},
  \citenamefont {Pablos},\ and\ \citenamefont
  {Rajagopal}}]{Casalderrey-Solana:2019ubu}%
  \BibitemOpen
  \bibfield  {author} {\bibinfo {author} {\bibfnamefont {J.}~\bibnamefont
  {Casalderrey-Solana}}, \bibinfo {author} {\bibfnamefont {G.}~\bibnamefont
  {Milhano}}, \bibinfo {author} {\bibfnamefont {D.}~\bibnamefont {Pablos}}, \
  and\ \bibinfo {author} {\bibfnamefont {K.}~\bibnamefont {Rajagopal}},\ }\href
  {\doibase 10.1007/JHEP01(2020)044} {\bibfield  {journal} {\bibinfo  {journal}
  {JHEP}\ }\textbf {\bibinfo {volume} {01}},\ \bibinfo {pages} {044} (\bibinfo
  {year} {2020})},\ \Eprint {http://arxiv.org/abs/1907.11248} {arXiv:1907.11248
  [hep-ph]} \BibitemShut {NoStop}%
\bibitem [{\citenamefont {Chang}\ \emph {et~al.}(2020)\citenamefont {Chang},
  \citenamefont {Tachibana},\ and\ \citenamefont {Qin}}]{Chang:2019sae}%
  \BibitemOpen
  \bibfield  {author} {\bibinfo {author} {\bibfnamefont {N.-B.}\ \bibnamefont
  {Chang}}, \bibinfo {author} {\bibfnamefont {Y.}~\bibnamefont {Tachibana}}, \
  and\ \bibinfo {author} {\bibfnamefont {G.-Y.}\ \bibnamefont {Qin}},\ }\href
  {\doibase 10.1016/j.physletb.2019.135181} {\bibfield  {journal} {\bibinfo
  {journal} {Phys. Lett. B}\ }\textbf {\bibinfo {volume} {801}},\ \bibinfo
  {pages} {135181} (\bibinfo {year} {2020})},\ \Eprint
  {http://arxiv.org/abs/1906.09562} {arXiv:1906.09562 [nucl-th]} \BibitemShut
  {NoStop}%
\bibitem [{\citenamefont {Du}\ \emph {et~al.}(2021)\citenamefont {Du},
  \citenamefont {Pablos},\ and\ \citenamefont {Tywoniuk}}]{Du:2020pmp}%
  \BibitemOpen
  \bibfield  {author} {\bibinfo {author} {\bibfnamefont {Y.-L.}\ \bibnamefont
  {Du}}, \bibinfo {author} {\bibfnamefont {D.}~\bibnamefont {Pablos}}, \ and\
  \bibinfo {author} {\bibfnamefont {K.}~\bibnamefont {Tywoniuk}},\ }\href
  {\doibase 10.1007/JHEP03(2021)206} {\bibfield  {journal} {\bibinfo  {journal}
  {JHEP}\ }\textbf {\bibinfo {volume} {03}},\ \bibinfo {pages} {206} (\bibinfo
  {year} {2021})},\ \bibinfo {note} {[JHEP21,206(2020)]},\ \Eprint
  {http://arxiv.org/abs/2012.07797} {arXiv:2012.07797 [hep-ph]} \BibitemShut
  {NoStop}%
\bibitem [{\citenamefont {Ke}\ and\ \citenamefont {Wang}(2021)}]{Ke:2020clc}%
  \BibitemOpen
  \bibfield  {author} {\bibinfo {author} {\bibfnamefont {W.}~\bibnamefont
  {Ke}}\ and\ \bibinfo {author} {\bibfnamefont {X.-N.}\ \bibnamefont {Wang}},\
  }\href {\doibase 10.1007/JHEP05(2021)041} {\bibfield  {journal} {\bibinfo
  {journal} {JHEP}\ }\textbf {\bibinfo {volume} {05}},\ \bibinfo {pages} {041}
  (\bibinfo {year} {2021})},\ \Eprint {http://arxiv.org/abs/2010.13680}
  {arXiv:2010.13680 [hep-ph]} \BibitemShut {NoStop}%
\bibitem [{\citenamefont {Vaidya}(2020)}]{Vaidya:2020lih}%
  \BibitemOpen
  \bibfield  {author} {\bibinfo {author} {\bibfnamefont {V.}~\bibnamefont
  {Vaidya}},\ }\href@noop {} {\  (\bibinfo {year} {2020})},\ \Eprint
  {http://arxiv.org/abs/2010.00028} {arXiv:2010.00028 [hep-ph]} \BibitemShut
  {NoStop}%
\bibitem [{\citenamefont {Sadofyev}\ \emph {et~al.}(2021)\citenamefont
  {Sadofyev}, \citenamefont {Sievert},\ and\ \citenamefont
  {Vitev}}]{Sadofyev:2021ohn}%
  \BibitemOpen
  \bibfield  {author} {\bibinfo {author} {\bibfnamefont {A.~V.}\ \bibnamefont
  {Sadofyev}}, \bibinfo {author} {\bibfnamefont {M.~D.}\ \bibnamefont
  {Sievert}}, \ and\ \bibinfo {author} {\bibfnamefont {I.}~\bibnamefont
  {Vitev}},\ }\href@noop {} {\  (\bibinfo {year} {2021})},\ \Eprint
  {http://arxiv.org/abs/2104.09513} {arXiv:2104.09513 [hep-ph]} \BibitemShut
  {NoStop}%
\bibitem [{\citenamefont {Chatrchyan}\ \emph {et~al.}(2012)\citenamefont
  {Chatrchyan} \emph {et~al.}}]{Chatrchyan:2012nia}%
  \BibitemOpen
  \bibfield  {author} {\bibinfo {author} {\bibfnamefont {S.}~\bibnamefont
  {Chatrchyan}} \emph {et~al.} (\bibinfo {collaboration} {CMS}),\ }\href
  {\doibase 10.1016/j.physletb.2012.04.058} {\bibfield  {journal} {\bibinfo
  {journal} {Phys. Lett. B}\ }\textbf {\bibinfo {volume} {712}},\ \bibinfo
  {pages} {176} (\bibinfo {year} {2012})},\ \Eprint
  {http://arxiv.org/abs/1202.5022} {arXiv:1202.5022 [nucl-ex]} \BibitemShut
  {NoStop}%
\bibitem [{\citenamefont {Sirunyan}\ \emph {et~al.}(2021)\citenamefont
  {Sirunyan} \emph {et~al.}}]{Sirunyan:2021jty}%
  \BibitemOpen
  \bibfield  {author} {\bibinfo {author} {\bibfnamefont {A.~M.}\ \bibnamefont
  {Sirunyan}} \emph {et~al.} (\bibinfo {collaboration} {CMS}),\ }\href
  {\doibase 10.1007/JHEP05(2021)116} {\bibfield  {journal} {\bibinfo  {journal}
  {JHEP}\ }\textbf {\bibinfo {volume} {05}},\ \bibinfo {pages} {116} (\bibinfo
  {year} {2021})},\ \Eprint {http://arxiv.org/abs/2101.04720} {arXiv:2101.04720
  [hep-ex]} \BibitemShut {NoStop}%
\bibitem [{\citenamefont {Karch}\ and\ \citenamefont
  {Katz}(2002)}]{Karch:2002sh}%
  \BibitemOpen
  \bibfield  {author} {\bibinfo {author} {\bibfnamefont {A.}~\bibnamefont
  {Karch}}\ and\ \bibinfo {author} {\bibfnamefont {E.}~\bibnamefont {Katz}},\
  }\href {\doibase 10.1088/1126-6708/2002/06/043} {\bibfield  {journal}
  {\bibinfo  {journal} {JHEP}\ }\textbf {\bibinfo {volume} {06}},\ \bibinfo
  {pages} {043} (\bibinfo {year} {2002})},\ \Eprint
  {http://arxiv.org/abs/hep-th/0205236} {arXiv:hep-th/0205236} \BibitemShut
  {NoStop}%
\bibitem [{\citenamefont {Chesler}\ and\ \citenamefont
  {Rajagopal}(2014)}]{Chesler:2014jva}%
  \BibitemOpen
  \bibfield  {author} {\bibinfo {author} {\bibfnamefont {P.~M.}\ \bibnamefont
  {Chesler}}\ and\ \bibinfo {author} {\bibfnamefont {K.}~\bibnamefont
  {Rajagopal}},\ }\href {\doibase 10.1103/PhysRevD.90.025033} {\bibfield
  {journal} {\bibinfo  {journal} {Phys. Rev. D}\ }\textbf {\bibinfo {volume}
  {90}},\ \bibinfo {pages} {025033} (\bibinfo {year} {2014})},\ \Eprint
  {http://arxiv.org/abs/1402.6756} {arXiv:1402.6756 [hep-th]} \BibitemShut
  {NoStop}%
\bibitem [{\citenamefont {Larkoski}\ \emph {et~al.}(2014)\citenamefont
  {Larkoski}, \citenamefont {Marzani}, \citenamefont {Soyez},\ and\
  \citenamefont {Thaler}}]{Larkoski:2014wba}%
  \BibitemOpen
  \bibfield  {author} {\bibinfo {author} {\bibfnamefont {A.~J.}\ \bibnamefont
  {Larkoski}}, \bibinfo {author} {\bibfnamefont {S.}~\bibnamefont {Marzani}},
  \bibinfo {author} {\bibfnamefont {G.}~\bibnamefont {Soyez}}, \ and\ \bibinfo
  {author} {\bibfnamefont {J.}~\bibnamefont {Thaler}},\ }\href {\doibase
  10.1007/JHEP05(2014)146} {\bibfield  {journal} {\bibinfo  {journal} {JHEP}\
  }\textbf {\bibinfo {volume} {05}},\ \bibinfo {pages} {146} (\bibinfo {year}
  {2014})},\ \Eprint {http://arxiv.org/abs/1402.2657} {arXiv:1402.2657
  [hep-ph]} \BibitemShut {NoStop}%
\bibitem [{\citenamefont {Khachatryan}\ \emph {et~al.}(2016)\citenamefont
  {Khachatryan} \emph {et~al.}}]{Khachatryan:2016tfj}%
  \BibitemOpen
  \bibfield  {author} {\bibinfo {author} {\bibfnamefont {V.}~\bibnamefont
  {Khachatryan}} \emph {et~al.} (\bibinfo {collaboration} {CMS}),\ }\href
  {\doibase 10.1007/JHEP11(2016)055} {\bibfield  {journal} {\bibinfo  {journal}
  {JHEP}\ }\textbf {\bibinfo {volume} {11}},\ \bibinfo {pages} {055} (\bibinfo
  {year} {2016})},\ \Eprint {http://arxiv.org/abs/1609.02466} {arXiv:1609.02466
  [nucl-ex]} \BibitemShut {NoStop}%
\bibitem [{\citenamefont {van~der Schee}(2013)}]{vanderSchee:2012qj}%
  \BibitemOpen
  \bibfield  {author} {\bibinfo {author} {\bibfnamefont {W.}~\bibnamefont
  {van~der Schee}},\ }\href {\doibase 10.1103/PhysRevD.87.061901} {\bibfield
  {journal} {\bibinfo  {journal} {Phys. Rev. D}\ }\textbf {\bibinfo {volume}
  {87}},\ \bibinfo {pages} {061901} (\bibinfo {year} {2013})},\ \Eprint
  {http://arxiv.org/abs/1211.2218} {arXiv:1211.2218 [hep-th]} \BibitemShut
  {NoStop}%
\bibitem [{\citenamefont {Habich}\ \emph {et~al.}(2015)\citenamefont {Habich},
  \citenamefont {Nagle},\ and\ \citenamefont {Romatschke}}]{Habich:2014jna}%
  \BibitemOpen
  \bibfield  {author} {\bibinfo {author} {\bibfnamefont {M.}~\bibnamefont
  {Habich}}, \bibinfo {author} {\bibfnamefont {J.~L.}\ \bibnamefont {Nagle}}, \
  and\ \bibinfo {author} {\bibfnamefont {P.}~\bibnamefont {Romatschke}},\
  }\href {\doibase 10.1140/epjc/s10052-014-3206-7} {\bibfield  {journal}
  {\bibinfo  {journal} {Eur. Phys. J. C}\ }\textbf {\bibinfo {volume} {75}},\
  \bibinfo {pages} {15} (\bibinfo {year} {2015})},\ \Eprint
  {http://arxiv.org/abs/1409.0040} {arXiv:1409.0040 [nucl-th]} \BibitemShut
  {NoStop}%
\bibitem [{\citenamefont {Lekaveckas}\ and\ \citenamefont
  {Rajagopal}(2014)}]{Lekaveckas:2013lha}%
  \BibitemOpen
  \bibfield  {author} {\bibinfo {author} {\bibfnamefont {M.}~\bibnamefont
  {Lekaveckas}}\ and\ \bibinfo {author} {\bibfnamefont {K.}~\bibnamefont
  {Rajagopal}},\ }\href {\doibase 10.1007/JHEP02(2014)068} {\bibfield
  {journal} {\bibinfo  {journal} {JHEP}\ }\textbf {\bibinfo {volume} {02}},\
  \bibinfo {pages} {068} (\bibinfo {year} {2014})},\ \Eprint
  {http://arxiv.org/abs/1311.5577} {arXiv:1311.5577 [hep-th]} \BibitemShut
  {NoStop}%
\bibitem [{\citenamefont {Hatta}\ \emph {et~al.}(2011)\citenamefont {Hatta},
  \citenamefont {Iancu}, \citenamefont {Mueller},\ and\ \citenamefont
  {Triantafyllopoulos}}]{Hatta:2011gh}%
  \BibitemOpen
  \bibfield  {author} {\bibinfo {author} {\bibfnamefont {Y.}~\bibnamefont
  {Hatta}}, \bibinfo {author} {\bibfnamefont {E.}~\bibnamefont {Iancu}},
  \bibinfo {author} {\bibfnamefont {A.~H.}\ \bibnamefont {Mueller}}, \ and\
  \bibinfo {author} {\bibfnamefont {D.~N.}\ \bibnamefont
  {Triantafyllopoulos}},\ }\href {\doibase 10.1016/j.nuclphysb.2011.04.011}
  {\bibfield  {journal} {\bibinfo  {journal} {Nucl. Phys. B}\ }\textbf
  {\bibinfo {volume} {850}},\ \bibinfo {pages} {31} (\bibinfo {year} {2011})},\
  \Eprint {http://arxiv.org/abs/1102.0232} {arXiv:1102.0232 [hep-th]}
  \BibitemShut {NoStop}%
\bibitem [{\citenamefont {Zapp}\ \emph {et~al.}(2009)\citenamefont {Zapp},
  \citenamefont {Ingelman}, \citenamefont {Rathsman}, \citenamefont {Stachel},\
  and\ \citenamefont {Wiedemann}}]{Zapp:2008gi}%
  \BibitemOpen
  \bibfield  {author} {\bibinfo {author} {\bibfnamefont {K.}~\bibnamefont
  {Zapp}}, \bibinfo {author} {\bibfnamefont {G.}~\bibnamefont {Ingelman}},
  \bibinfo {author} {\bibfnamefont {J.}~\bibnamefont {Rathsman}}, \bibinfo
  {author} {\bibfnamefont {J.}~\bibnamefont {Stachel}}, \ and\ \bibinfo
  {author} {\bibfnamefont {U.~A.}\ \bibnamefont {Wiedemann}},\ }\href {\doibase
  10.1140/epjc/s10052-009-0941-2} {\bibfield  {journal} {\bibinfo  {journal}
  {Eur. Phys. J. C}\ }\textbf {\bibinfo {volume} {60}},\ \bibinfo {pages} {617}
  (\bibinfo {year} {2009})},\ \Eprint {http://arxiv.org/abs/0804.3568}
  {arXiv:0804.3568 [hep-ph]} \BibitemShut {NoStop}%
\bibitem [{\citenamefont {Mehtar-Tani}\ \emph {et~al.}(2012)\citenamefont
  {Mehtar-Tani}, \citenamefont {Salgado},\ and\ \citenamefont
  {Tywoniuk}}]{MehtarTani:2011tz}%
  \BibitemOpen
  \bibfield  {author} {\bibinfo {author} {\bibfnamefont {Y.}~\bibnamefont
  {Mehtar-Tani}}, \bibinfo {author} {\bibfnamefont {C.~A.}\ \bibnamefont
  {Salgado}}, \ and\ \bibinfo {author} {\bibfnamefont {K.}~\bibnamefont
  {Tywoniuk}},\ }\href {\doibase 10.1016/j.physletb.2011.12.042} {\bibfield
  {journal} {\bibinfo  {journal} {Phys. Lett. B}\ }\textbf {\bibinfo {volume}
  {707}},\ \bibinfo {pages} {156} (\bibinfo {year} {2012})},\ \Eprint
  {http://arxiv.org/abs/1102.4317} {arXiv:1102.4317 [hep-ph]} \BibitemShut
  {NoStop}%
\bibitem [{\citenamefont {Hulcher}\ \emph {et~al.}(2018)\citenamefont
  {Hulcher}, \citenamefont {Pablos},\ and\ \citenamefont
  {Rajagopal}}]{Hulcher:2017cpt}%
  \BibitemOpen
  \bibfield  {author} {\bibinfo {author} {\bibfnamefont {Z.}~\bibnamefont
  {Hulcher}}, \bibinfo {author} {\bibfnamefont {D.}~\bibnamefont {Pablos}}, \
  and\ \bibinfo {author} {\bibfnamefont {K.}~\bibnamefont {Rajagopal}},\ }\href
  {\doibase 10.1007/JHEP03(2018)010} {\bibfield  {journal} {\bibinfo  {journal}
  {JHEP}\ }\textbf {\bibinfo {volume} {03}},\ \bibinfo {pages} {010} (\bibinfo
  {year} {2018})},\ \Eprint {http://arxiv.org/abs/1707.05245} {arXiv:1707.05245
  [hep-ph]} \BibitemShut {NoStop}%
\bibitem [{\citenamefont {He}\ \emph {et~al.}(2012)\citenamefont {He},
  \citenamefont {Vitev},\ and\ \citenamefont {Zhang}}]{He:2011pd}%
  \BibitemOpen
  \bibfield  {author} {\bibinfo {author} {\bibfnamefont {Y.}~\bibnamefont
  {He}}, \bibinfo {author} {\bibfnamefont {I.}~\bibnamefont {Vitev}}, \ and\
  \bibinfo {author} {\bibfnamefont {B.-W.}\ \bibnamefont {Zhang}},\ }\href
  {\doibase 10.1016/j.physletb.2012.05.054} {\bibfield  {journal} {\bibinfo
  {journal} {Phys. Lett. B}\ }\textbf {\bibinfo {volume} {713}},\ \bibinfo
  {pages} {224} (\bibinfo {year} {2012})},\ \Eprint
  {http://arxiv.org/abs/1105.2566} {arXiv:1105.2566 [hep-ph]} \BibitemShut
  {NoStop}%
\bibitem [{Note1()}]{Note1}%
  \BibitemOpen
  \bibinfo {note} {Note that our dijets follow the measured dijet asymmetry
  distribution and that transverse momentum is hence not balanced.}\BibitemShut
  {Stop}%
\bibitem [{\citenamefont {Cao}\ \emph {et~al.}(2017)\citenamefont {Cao} \emph
  {et~al.}}]{Cao:2017zih}%
  \BibitemOpen
  \bibfield  {author} {\bibinfo {author} {\bibfnamefont {S.}~\bibnamefont
  {Cao}} \emph {et~al.} (\bibinfo {collaboration} {JETSCAPE}),\ }\href
  {\doibase 10.1103/PhysRevC.96.024909} {\bibfield  {journal} {\bibinfo
  {journal} {Phys. Rev. C}\ }\textbf {\bibinfo {volume} {96}},\ \bibinfo
  {pages} {024909} (\bibinfo {year} {2017})},\ \Eprint
  {http://arxiv.org/abs/1705.00050} {arXiv:1705.00050 [nucl-th]} \BibitemShut
  {NoStop}%
\bibitem [{\citenamefont {Sirunyan}\ \emph {et~al.}(2018)\citenamefont
  {Sirunyan} \emph {et~al.}}]{Sirunyan:2018jqr}%
  \BibitemOpen
  \bibfield  {author} {\bibinfo {author} {\bibfnamefont {A.~M.}\ \bibnamefont
  {Sirunyan}} \emph {et~al.} (\bibinfo {collaboration} {CMS}),\ }\href
  {\doibase 10.1007/JHEP05(2018)006} {\bibfield  {journal} {\bibinfo  {journal}
  {JHEP}\ }\textbf {\bibinfo {volume} {05}},\ \bibinfo {pages} {006} (\bibinfo
  {year} {2018})},\ \Eprint {http://arxiv.org/abs/1803.00042} {arXiv:1803.00042
  [nucl-ex]} \BibitemShut {NoStop}%
\end{thebibliography}%

\end{document}